\documentclass{cpcauth}

\usepackage{epsfig}

\usepackage{amssymb}
\let\boldsymbol\pol
\begin{document}

\begin{frontmatter}



\title{Dynamic Monte Carlo Simulations for a Square-Lattice 
Ising Ferromagnet with a Phonon Heat Bath}


\author{Kyungwha Park}
\address{School of Computational Science and Information Technology,
Florida State University, Tallahassee, Florida 32306, USA}

\author{M.~A.\ Novotny}
\address{Department of Physics and Astronomy, Mississippi 
State University, Mississippi State, Mississippi 39762, USA}

\begin{abstract}
We derive a direct connection between Monte Carlo time and
physical time in terms of physical parameters, 
using a quantum Hamiltonian with a $d$-dimensional
{\it phonon} heat bath interacting with a square-lattice Ising ferromagnet. 
Based on the calculated transition rates, we perform dynamic 
Monte Carlo simulations using absorbing Markov chains to measure 
the lifetimes of the metastable state at low temperatures. We 
also calculate the lifetimes analytically using absorbing Markov chains.
The phonon dynamic gives field-dependent prefactors in the lifetimes 
at low temperatures, that are different from the piecewise field-independent 
prefactors obtained from the Glauber dynamic. 
\end{abstract}

\end{frontmatter}

\section{Introduction}

Dynamics of classical Ising spin systems have been studied
extensively using the Glauber dynamic \cite{GLAU}.
The Glauber dynamic was later also derived by Martin \cite{MART} 
starting from a quantum Hamiltonian
consisting of an exchange interaction and a coupling to a {\it fermionic}
thermal heat bath attached to each spin. This derivation was
performed in a weak-coupling limit
in which the square of the coupling constant multiplied by a 
characteristic time becomes an arbitrary finite constant.
In this limit, the time evolution of the system was proved to be Markovian.

A different dynamic would change the time evolution of the system, but not
its equilibrium properties. Here we study the quantum nearest-neighbor Ising
ferromagnet with an applied longitudinal magnetic field and a linear
coupling between the Ising spins and a phonon (i.e. bosonic) heat bath.
We calculated the transition rates from one configuration to
another resulting from the spin-phonon coupling. We then applied
these transition rates to dynamic Monte Carlo simulations
and measured the average lifetimes $\langle \tau \rangle$ of the metastable 
state \cite{NOVO01}. To measure $\langle \tau \rangle$, 
we set the initial configuration with all
spins up and applied a magnetic field $H $$<$$ 0$. We define
$\langle \tau \rangle$ as the number of spin-flip attempts 
until the magnetization reaches zero.
Exact predictions \cite{JORD} of the lifetime $\langle \tau \rangle$
as $T $$\rightarrow$$ 0$ are given by
\begin{equation}
T {\mathrm {ln}} \langle \tau \rangle\!=\!
\Gamma(H,J)\!=\!8J\ell_c - 2 |H| (\ell_c^2 - \ell_c + 1) \;,
\label{eq:taulowT}
\end{equation}
where the linear critical droplet size is 
$\ell_c $$=$$ \lceil 2J/|H| \rceil$,
$\lceil x \rceil$ denotes the smallest integer not less than
$x$, and $\Gamma$ is the energy cost of a critical droplet. 
The critical droplet is a cluster of overturned spins which is 
an $\ell_c \times (\ell_c -1)$ rectangle with one additional overturned
spin on one of the long sides of the rectangle.
This formula is valid for $2J/|H|$ not an integer and $|H| $$<$$ 4 J$.
Here we show that the specific dynamic crucially affects the time evolution
of the system.

\section{Dynamic Quantum Model}

The total Hamiltonian we use is
${\mathcal H}={\mathcal H}_{\mathrm{sp}}+{\mathcal H}_{\mathrm{ph}}
+{\mathcal H}_{\mathrm {sp-ph}}$.
The spin Hamiltonian ${\mathcal H}_{\mathrm {sp}}$ and phonon
Hamiltonian ${\mathcal H}_{\mathrm {ph}}$ are given by
\begin{eqnarray}
{\mathcal H}_{\mathrm {sp}} &=& -J \sum_{\langle i,j \rangle} 
\sigma_i^z \sigma_j^z
-H_z \sum_{i} \sigma_i^z \\
{\mathcal H}_{\mathrm {ph}}&=& 
\sum_{\vec{q}} \hbar \omega_{\vec{q}} c^{\dagger}_{\vec{q}}
c_{\vec{q}} ~,
\end{eqnarray}
where the first summation runs over nearest-neighbor sites only.
$J($$>$$0)$ is the exchange coupling constant, $\sigma_j^z$
are the $z$ components of Pauli spin operators attached to
lattice site $j$, $H_z$ is a longitudinal magnetic field,
$\vec{q}$ is the wave vector of a phonon mode, $\omega_{\vec{q}}$ is 
an angular frequency of the phonon mode with $\vec{q}$, and
$c^{\dagger}_{\vec{q}}$ and $c_{\vec{q}}$ are creation and annihilation 
operators of the phonon mode with $\vec{q}$.
For simplicity, we ignore the direction of $\vec{q}$,
so hereafter we drop the vector symbol on $\vec{q}$.
If we consider only linear coupling between
spin operators $V(\vec{\sigma})$ and strains caused by
phonons, the general spin-phonon
interaction for a single spin $\vec{\sigma}$ at the origin in 
real space is $\sum_{q} \sqrt{\frac{\hbar}{2NM\omega_{q}}}
(iqV(\vec{\sigma})c^{\dagger}_{q} 
- iqV^{\dagger}(\vec{\sigma})c_{q})$,
where $N$ is the number of unit cells consisting of one spin, and $M$ is 
the mass of the unit cell.  When the Ising spins are located at positions
$\vec{R}_j$, the spin-phonon interaction Hamiltonian \cite{HART} is:
\begin{equation}
\!\!{\mathcal H}_{\mathrm {sp-ph}}\!=\! \lambda \! \sum_{j,q}
\!\! \sqrt{\!\frac{\hbar}{2NM\omega_{q}}} (iq~ \sigma_j^x) 
(c^{\dagger}_{q} - c_{q}) e^{i \vec{q} \cdot \vec{R}_j}, 
\end{equation}
where $\lambda$ is the coupling strength between the spin
system and the phonon heat bath.   

With the given spin Hamiltonian, the dynamic
is determined by the generalized master equation \cite{BLUM,LEUE}:
\begin{eqnarray}
\! \! \frac{{\mathrm d}\rho(t)_{m^{\prime} m}}{{\mathrm d}t}&=&\frac{i}{\hbar}
[\rho(t),{\mathcal H}_{\mathrm {sp}}]_{m^{\prime} m} 
+ \delta_{m^{\prime} m} \sum_{n \neq m} \rho(t)_{nn} W_{mn} \nonumber \\
& &- \gamma_{m^{\prime} m} \rho(t)_{m^{\prime} m}~, 
\nonumber \\
\gamma_{m^{\prime} m}&=&\frac{W_m+W_{m^{\prime}}}{2}, \; \;
W_m = \sum_{k \neq m} W_{km},
\end{eqnarray}
where $\rho(t)$ is the time dependent density matrix of the 
spin system, $m^{\prime}$, $n$, $k$, and $m$ are eigenstates of 
${\mathcal H}_{\mathrm{sp}}$, 
$\rho(t)_{m^{\prime} m}$$=$$\langle m^{\prime}| \rho(t) | m \rangle$,
and $W_{km}$ is a transition rate from the $m$-th to the $k$-th eigenstate. 
In our case, because there are no off-diagonal terms in 
${\mathcal H}_{\mathrm{sp}}$, this generalized master equation 
becomes identical
to the master equation which Glauber used in his paper \cite{GLAU},
but with different $W_{km}$.
Assuming that the correlation time of the heat bath is much shorter
than the times of interest, we integrate over all degrees of freedom
of the heat bath in order to obtain the transition rates.
The transition rate from the $l$-th to the $k$-th eigenstate 
of ${\mathcal H}_{\mathrm{sp}}$ becomes
\begin{eqnarray}
W_{k,l}\!&=&\! \frac{2\pi}{\hbar} \! \sum_{q,n_{q}} \!
\left|{ \langle n_{q} + 1, 
k ~|{\mathcal H}_{\mathrm {sp-ph}}|~n_{q},l \rangle }
\right|^2 \nonumber \\
& & \! \times \langle n_{q} |\rho_{\mathrm{ph}}|n_{q} \rangle 
\delta(E_l - E_k - \hbar \omega_{q}) \;,
\end{eqnarray}
with the energy eigenvalues of ${\mathcal H}_{\mathrm {sp}}$ 
$E_l $$>$$ E_k$.  Here
$n_q$ is the average occupation number of the phonon mode with 
$q$, and $\rho_{\mathrm{ph}}$ is the density matrix of the phonon
bath.  We can calculate the transition rate when $E_l $$<$$ E_k$ similarly.  
Eventually we obtain
\begin{eqnarray}
W_{k,l} &=& \frac{\lambda^2}{\Theta \eta \hbar^{d+1} c^{d+2}} \left|{
\frac{(E_k-E_l)^d}{e^{\beta(E_k-E_l)}-1} }\right|~,
\end{eqnarray}
where $d$ is the dimension of the heat bath, 
$\Theta=2$ ($2\pi$) for $d$$=$1,2 (3), 
$\eta$ is the mass density of a unit cell, $c$ is the sound
velocity, and $\beta$$=$$1/k_B T$.
The two major differences from the Glauber dynamic are
the energy term in the numerator and the negative
sign in the denominator. In the limit $T$$\rightarrow$0, 
the transition rates vanish when $E_l$$=$$E_k$ (this can occur for
$|H|$$=$$2J, 4J$). The transition rates satisfy detailed balance.

\section{Lifetimes of the Metastable State}

\begin{center}
\begin{figure}[t]
\includegraphics[width=7cm,height=5cm]{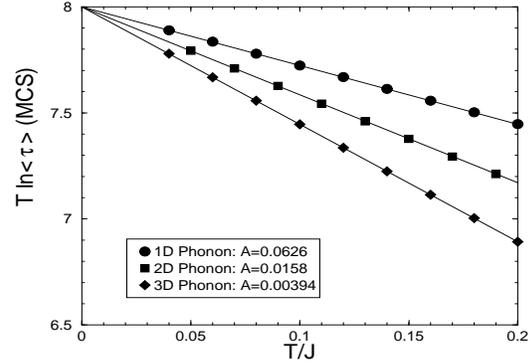}
\caption{$T {\mathrm{ln}} \langle \tau \rangle$ in units of
Monte Carlo steps as a function of $T$ at $|H|$$=$$2J$
with the $d$-dimensional phonon dynamic. The symbols are 
our $s$$=$2 MCAMC data. The solid lines denote extrapolations
to zero temperature.
The prefactors $A$ can be read from the slopes of the lines.}
\label{eps1}
\end{figure}
\end{center}

We perform dynamic Monte Carlo simulations using Absorbing Markov Chains 
(MCAMC) \cite{NOVO01,NOVO97} at low temperatures with both the Glauber 
and $d$-dimensional phonon dynamics. 
For $2J $$<$$ |H| $$<$$ 4J$, the critical droplet consists of a single 
overturned spin, so the $n$-fold way algorithm \cite{BORT} 
($s$$=$$1$ MCAMC) gives adequate speed-ups. However, 
when $|H|$ approaches 2 from below or above, the $s$$=$$2$ MCAMC
algorithm is needed to prevent the system from fluctuating 
between the all-spins-up state and the state with all spins up 
except for one overturned spin. For $J $$<$$ |H| $$<$$ 2J$, 
the critical droplet has an L-shape formed by three overturned spins.  
This necessitates using $s$$=$$3$ MCAMC to prevent the system from 
fluctuating between the state with all spins up except for one 
overturned spin and the state with two overturned nearest neighbor spins.
Periodic boundary conditions are used. For both dynamics, 
average lifetimes are measured over 2000 escapes with the system 
size $L$=24. The range of temperatures used is between $T/J$=0.04
and $T/J$=0.2. At very low temperatures for a given field, 
MPFUN \cite{BAIL} is used for high-precision calculations.  
Since at low temperatures the lifetime is the inverse of the probability 
that the system escapes from the metastable well,
the lifetime is written as
\begin{eqnarray}
\langle \tau \rangle &=& A e^{\beta \Gamma}~,
\end{eqnarray}
where $A$ is a prefactor and $\Gamma$ is given in Eq.(\ref{eq:taulowT}).  
In a given field, the prefactor $A$ and $\Gamma$ can be extrapolated to zero
temperature from the measured lifetimes
: $T\ln \langle \tau \rangle = T \ln A + \Gamma$ (Figure \ref{eps1}).

\begin{center}
\begin{figure}[t]
\includegraphics[width=7cm]{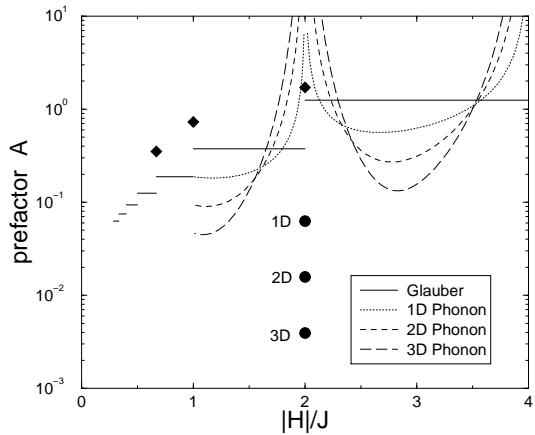}
\caption{Prefactors $A$ from the $d$-dimensional phonon and 
Glauber dynamics as functions of $|H|$ as $T \rightarrow 0$.  
Filled circles (diamonds) denote the prefactors from the phonon
(Glauber) dynamic when $2J/|H|$ is an integer.
When $2J/|H|$ is not an integer, the prefactors can be calculated 
using absorbing Markov chains.}
\label{eps2}
\end{figure}
\end{center}

The exact lifetimes can be analytically 
obtained from Absorbing Markov Chains (AMC) 
\cite{NOVO01,NOVO97} in principle. For 2$J$$<$$|H|$$<$4$J$, $\ell_c$=1, 
two transient states are needed while for $J$$<$$|H|$$<$2$J$, $\ell_c$=2, 
seven transient states are needed. 
For fields lower than $J$, a large number of transient states are
needed so practically it is not possible to compute the exact lifetimes
from AMC. The Glauber and phonon dynamics provide the same
energy barrier $\Gamma$. However, the prefactor $A$ in the lifetime
resulting from the phonon dynamic is different 
from that from the Glauber dynamic.
The Glauber prefactor is known to be $A$=5/4 for $\ell_c$=1, 
$A$=3/8 for $\ell_c$=2 \cite{NOVO97}, and 
$3/(8 (\ell_c-1))$ for $\ell_c $$\geq$$ 2$ 
\cite{BOVI}. The $d$-dimensional phonon dynamic gives a prefactor $A$ that 
has a non-constant derivative with respect to $|H|$ and that
depends on the dimension of the heat bath. 
For $2J $$<$$ |H| $$<$$ 4J$, 
\begin{equation}
A(|H|,d)\!=\!
\frac{4(2|H|-4J)^d + (8J-2|H|)^d}{4(2|H|-4J)^d (8J-2|H|)^d}\;.
\label{eq:a1}
\end{equation}
In the limit that $d$$\rightarrow$$0$, $A$ approaches the
Glauber prefactor, 5/4. For an infinite-dimensional bath, 
$A$$\rightarrow$$\infty$ for 2$J$$<$$|H|$$<$5$J$/2
and 7$J$/2$<$$|H|$$<$4$J$, $A$$\rightarrow$1/4 for $|H|$$=$5$J$/2, 
$A$$\rightarrow$0 for 5$J$/2$<$$|H|$$<$7$J$/2, and 
$A$$\rightarrow$$1$ for $|H|$$=$7$J$/2.
For $J $$<$$ |H| $$<$$ 2J$, 
\begin{equation}
A(|H|,d)=
\frac{|H|^d + 2(2J-|H|)^d}{2^{d+3}|H|^d (2J-|H|)^d}\;.
\label{eq:a2}
\end{equation}
In the limit that $d \rightarrow 0$, $A$ approaches the
Glauber prefactor, 3/8. For an infinite dimensional bath,
$A \rightarrow 0$ for $J$$<$$|H|$$<$3$J$/2, 
$A \rightarrow 1/8$ for $|H|$$=$3$J$/2, and
$A \rightarrow \infty$ for 3$J$/2$<$$|H|$$<$2$J$.
The expressions (\ref{eq:a1}) and (\ref{eq:a2}) for the prefactor
are exact as $T \rightarrow 0$. At nonzero temperatures 
there are correction terms on the order of $e^{-\beta \delta}$ 
where $\delta$ depends on the field. 
Figure \ref{eps2} shows the prefactor $A$ vs. $|H|$ for 
the phonon and Glauber dynamics.
The prefactors for integer values of $2J/|H|$ are 
obtained from the MCAMC simulation data, while the prefactors
in other magnetic fields are from Eqs.
(\ref{eq:a1}) and (\ref{eq:a2}). 
We have confirmed that the measured prefactors agree with the calculated
values for the $d$-dimensional phonon dynamic.

The prefactor from the phonon dynamic diverges as $|H| $$\rightarrow$$ 2J$ 
or 4$J$ because certain spin flips are not allowed. For $|H|$$=$4$J$, 
the probability to flip a spin in the all-spin-up state vanishes, 
while for $|H|$$=$2$J$ the probability to flip a spin in the configuration
with three nearest-neighbor spins up and one down vanishes.
At $|H|$$=$2$J$, the energy barrier is 2(8$J$$-$2$|H|$) for the phonon
dynamic because the system reaches a critical droplet by creating
two overturned 2nd-neighbor spins or two overturned 3rd-neighbor spins. 
Figure \ref{eps1} shows $\Gamma$$=$8$J$ at $|H|$$=$2$J$.
However, this behaviour does not occur at $|H|$$=$$J$, so the prefactor 
from the phonon dynamic is finite as $|H|$ approaches $J$.
For bulk iron, the exchange coupling constant $J$ corresponds to
about 300~T \cite{BROW}. So $|H|$$=$$J$ is too high to 
achieve in a laboratory although the actual value of $J$ may 
vary with structure and surrounding environment of materials.

\section{Conclusion}

We derived a relationship between Monte Carlo time and physical time
in terms of material parameters, starting from a quantum Hamiltonian 
with a phonon heat bath. We applied
this dynamic to the square-lattice Ising ferromagnet and measured the
lifetime of the metastable state. This dynamic gives a very different
low-temperature prefactor from the Glauber dynamic. \\
\\
Partially funded by NSF DMR-9871455.

\end{document}